\documentclass[twocol]{ametsocV6.1}
\usepackage{physics}
\usepackage{graphicx,nicefrac}

\renewcommand*{\vec}[1]{\mathbf{#1}}
\newcommand{\dss}{\displaystyle}
\newcommand{\ordr}[1]{^{(#1)}}
\newcommand{\vecu}{\vb{u}}

\title{Scaling approaches to quasi-geostrophic theory for moist, precipitating air}

\authors{Daniel B\"{a}umer,\aff{a}\correspondingauthor{Daniel B\"{a}umer, daniel.baeumer@univie.ac.at}
Sabine Hittmeir,\aff{a}
Rupert Klein,\aff{b}
}

\affiliation{\aff{a}{Fakult\"{a}t f\"{u}r Mathematik, Universit\"{a}t Wien, Vienna, Austria}\\
\aff{b}{FB Mathematik und Informatik, Freie Universit\"{a}t Berlin, Berlin, Germany}
}

\abstract{Quasi-geostrophic (QG) theory is of fundamental importance in the study of large-scale atmospheric flows. In recent years, there has been growing interest in extending the classical QG plus Ekman friction layer model (QG-Ekman) to systematically include additional physical processes known to significantly contribute to real-life weather phenomena. This paper lays the foundation for combining two of these developments, namely Smith and Stechmann's family of \emph{Precipitating Quasi-Geostrophic} (PQG) models (J.\ Atmos.\ Sci, {\bfseries 74}, 3285--3303, 2017) on the one hand, and the extension of QG-Ekman for dry air by a strongly \emph{Diabatic Layer} (DL) of intermediate height (QG-DL-Ekman) in (J.\ Atmos.\ Sci, {\bfseries 79}, 887--905, 2022) on the other hand. To this end, Smith and Stechmann's PQG equations for sound-proof motions are first corroborated within a general asymptotic modeling framework starting from a full compressible flow model. The derivations show that the PQG model family is naturally embedded in the asymptotic hierarchy of scale-dependent atmospheric flow models introduced by one of the present authors in (Ann.\ Rev.\ Fluid Mech., {\bfseries 42}, 249--274). Particular emphasis is then placed on an asymptotic scaling regime for PQG that accounts for a generic Kessler-type bulk microphysics closure and is compatible with QG-DL-Ekman theory. The detailed derivation of a moist QG-DL-Ekman model is deferred to a future publication.}

\begin{document}

\maketitle


\section{Introduction}

Ever since its inception, the mathematical model provided by quasi-geostrophic (QG) theory has proven highly successful as a streamlined setting for the explanation of major features of large-scale atmospheric flow in the midlatitudes. Its textbook derivation by scale analysis and asymptotic expansion, as found, e.g., in \citep{pedlosky1987}, is a beautiful example of the interplay between theoretical meteorology and applied mathematics. 

This well-established model, however, does not describe the contributions of moisture to the large-scale flow in an explicit fashion: only balance equations for dry air are prescribed, and the only way to integrate the vitally important effect of latent heat on the energy budget without extending the model itself is its parametrization as a heat source in the temperature equation. This approach has been pursued, e.g., by \citet{devries2010} who also provide a discussion of different parametrization schemes.

As far as actual extensions of the dry QG model are concerned, they still tend to treat moisture as a supplement; the model proposed by \citet{lapeyre2004} is emblematic of this approach, in which the authors take QG for dry air as the starting point and then formulate an equation for the mixing ratio of water vapor based on certain ad hoc assumptions. Steps toward a more systematic embedding of moisture into QG theory were made by \citet{monteiro2016}. These authors \emph{did} include moisture in their scaling and asymptotic analysis, but their derivation took place within the confines of a one-layer shallow-water model.

Seeking a more generally valid extension of QG theory that supplies balance equations for water vapor and precipitation and explicitly models the impact of moisture on the energy budget therefore poses an interesting challenge. Moreover, the mathematical derivation of such a model should proceed along the same lines as that of the classical theory. To this date, to the best of the authors' knowledge, only one model has been proposed that meets these criteria: the \emph{precipitating quasi-geostrophic} (PQG) equations of \citet{smith2017}. To clarify their key ideas, the authors first provided a detailed derivation starting from a somewhat simplified cloud-resolving model, the \emph{fast autoconversion and rain evaporation} (FARE) system of \citet{hernandez2013}. That model adopts the Boussinesq equations for the dynamics, linearized thermodynamic relations, and the limit of fast conversion of cloud into rain water. The authors point out, however, that the PQG equations form an entire model family parameterized by the asymptotic scalings assumed for the adopted flow models, e.g., Boussinesq or anelastic, the thermodynamic equations of state, and the bulk microphysics closure (see sections~6 and 9 of their paper).

Common to the QG and PQG models is the assumption of a strong background stratification of the (equivalent) potential temperature which, true to the well-known QG scalings, is equivalent to a small (moist) internal wave Froude number. In this context, let us quote from \citet{smith2017}: ``... locally in some regions, such as in the vicinity of fronts, assumptions of strong moist stratification and/or classical QG scaling may not hold.'' Flow regimes of this kind are addressed explicitly by \citet{klein2022} through their QG-DL-Ekman triple-deck boundary layer theory. Within its additional diabatic layer (DL) of intermediate height of $\sim\! 3\, \text{km}$, the potential temperature is not restricted to small deviations from a given background stratification but can freely evolve, even towards neutral stratification, instead. This way, systematically stronger diabatic effects are allowed for in the DL than those accounted for in the classical (P)QG models. The QG-DL-Ekman theory has thus far been derived only for dry air flows, though, and within our current ongoing work we aim to extend it to include moist processes.

The present paper is our first step in this direction and it provides two main contributions: In sections~4 and 5 we illuminate the relationship between the PQG equations in their anelastic form (based on the FARE model for moist processes) on the one hand, and the system for moist compressible flow with the generic Kessler-type bulk microphysics closure of \citet{hittmeir2018} on the other hand. To this end, we proceed by systematically deriving the former from the latter utilizing asymptotic techniques. This will also demonstrate how the PQG model family can be embedded naturally in the rich hierarchy of known scale-dependent atmospheric models as discussed in \citep{klein2010}.

This paper's second main contribution is a self-contained derivation of a \emph{a particular version} of PQG-type equations that is set up for the subsequent inclusion in a moist QG-DL-Ekman triple-deck theory. The underlying scalings combine features of several of the aforementioned versions of PQG, which is why we place particular emphasis on the justification of and the reasoning behind our scaling choices. In section~6, this leads indeed to a PQG model that is tailor-made to be coupled with the diabatic layer of \citet{klein2022}. We will further explore its implications in future publications. To provide some first insight into the model's characteristics, we discuss in section~7 its related ``omega equation'', which often serves to diagnose upward vertical velocities in weather forecast model output.

Finally, the appendix sketches a derivation of the moist anelastic system of \citet{hernandez2013}. There, we also point to the key differences between an asymptotic analysis that starts from a fully compressible system and one that is based on the anelastic approximation. This, in particular, serves to explain the occurrence of additional background terms in our derivation of the buoyancy in section~5.


\section{The governing equations}

Our point of departure is a system that not only accurately describes compressible flow, but also includes fairly detailed moist thermodynamics (see \citet{cotton2011} for a broad discussion of possible modelling approaches). This model formulation goes back to \citet{hittmeir2018}, and it includes established bulk microphysics closures as proposed and investigated in \citep{kessler1995}, \citep{grabowski1996} and \citep{klein2006}:
\begin{subequations}\label{eq:GoverningEquations}
\begin{align}
\label{eq:GoverningEquations-a}
D_t\vec{u}+2(\vec{\Omega}\times\vec{v})_\parallel+\frac{1}{\rho}\grad_\parallel{p}
  & = 0, 
    \\
\label{eq:GoverningEquations-b}
D_tw+2(\vec{\Omega}\times\vec{v})_\perp+\frac{1}{\rho}\partial_zp
  & = -g, 
    \\
\label{eq:GoverningEquations-c}
D_t\rho_d+\rho_d(\div{\vec{v}})
  & = 0, 
    \\
\label{eq:GoverningEquations-d}
C D_t\ln\theta+\Sigma D_t\ln p+\frac{L(T)}{T}D_tq_v &= \nonumber \\
 c_lV_rq_r(\partial_z\ln\theta&+\frac{R_d}{c_{\text{pd}}}\partial_z\ln p), 
    \\
\label{eq:GoverningEquations-e}
D_tq_v
  & = S_{\text{ev}}-S_{\text{cd}}, 
    \\
\label{eq:GoverningEquations-f}
D_tq_c
  & = S_{\text{cd}}-S_{\text{cr}}-S_{\text{ac}}, 
    \\
\label{eq:GoverningEquations-g}
D_tq_r-\frac{1}{\rho_d}\partial_z(\rho_dV_rq_r)
  & = S_{\text{cr}}+S_{\text{ac}}-S_{\text{ev}},
\end{align}
\end{subequations}
with the additional relations
\begin{subequations}\label{eq:GoverningEquationsSupplement}
\begin{align}
\label{eq:GoverningEquationsSupplement-a}
\rho
  & =\rho_d(1+q_v+q_c+q_r), 
    \\
\label{eq:GoverningEquationsSupplement-b}
p
  & = R_d\rho_dT(1+\frac{q_v}{\frac{R_d}{R_v}}), 
    \\
\label{eq:GoverningEquationsSupplement-c}
\theta
  & = T\left(\frac{p_{\text{ref}}}{p}\right)^{\frac{\gamma-1}{\gamma}}, 
    \\
\label{eq:GoverningEquationsSupplement-d}
\vec{v}
  & = \vec{u}+w\vec{k}, 
    \\
\label{eq:GoverningEquationsSupplement-e}
S_{\text{ev}}
  & = C_{\text{ev}}\frac{p}{\rho}(q_{\text{vs}}-q_v)^+q_r, 
    \\
\label{eq:GoverningEquationsSupplement-f}
S_{\text{cd}}
  & = C_{\text{cn}}(q_v-q_{\text{vs}})^+q_{\text{cn}}+C_{\text{cd}}(q_v-q_{\text{vs}})q_c, 
    \\
\label{eq:GoverningEquationsSupplement-g}
S_{\text{ac}}
  & = C_{\text{ac}}(q_c-q_{\text{ac}})^+, 
    \\
\label{eq:GoverningEquationsSupplement-h}
S_{\text{cr}}
  & = C_{\text{cr}}q_cq_r, 
    \\
\label{eq:GoverningEquationsSupplement-i}
C 
  & = c_{\text{pd}}+c_{\text{pv}}q_v+c_l(q_c+q_r), 
    \\
\label{eq:GoverningEquationsSupplement-j}
\Sigma
  & = (\frac{c_{\text{pv}}}{c_{\text{pd}}}R_d-R_v)q_v+\frac{c_l}{c_{\text{pd}}}R_d(q_c+q_r), 
    \\
\label{eq:GoverningEquationsSupplement-k}
L(T) 
  & = L_{\text{ref}}+(c_{\text{pv}}-c_l)(T-T_{\text{ref}})\equiv L_{\text{ref}}\phi(T).
\end{align}
\end{subequations}
In the above equations, $(\vec{u}=(u,v,0),w,\rho,\rho_d,T,\theta,p,q_v,q_c,q_r,q_{\text{vs}})$ denote horizontal and vertical velocity, density, dry air density, temperature, potential temperature, pressure and the mixing ratios of water vapor, cloud water and rain, as well as the saturation mixing ratio, respectively; $g$ is the gravitational acceleration, $\Omega$ the earth rotation vector, and the subscripts $\parallel$ and $\perp$ indicate horizontal and vertical components, respectively. We denote the positive part of a function $f$ by $f^+$. As usual, $c_{\text{pd}}$ and $c_{\text{pv}}$ denote the specific heat capacities at constant pressure of dry air and water vapor, while $c_l$ is the heat capacity of liquid water, here assumed constant for simplicity, and $V_r$ is the terminal rainfall velocity; $R_d$ and $R_v$ are the gas constants for dry air and water vapor, $\gamma=c_{\text{pd}}\slash(c_{\text{pd}}-R_d)$ is the isentropic exponent of dry air, $\vec{k}$ the vertical unit vector, $p_{\text{ref}}=10^5\text{ Pa}$ the reference pressure and the material derivative is given by
\begin{equation}
D_t=\partial_t+\vec{v}\cdot\grad=\partial_t+\vec{u}\cdot\grad_\parallel+w\partial_z.
\end{equation}

In line with usual assumptions \citep{cotton2011}, the dry air mass obeys the continuity equation~\eqref{eq:GoverningEquations-c}. Note that the density appearing in the momentum equations \eqref{eq:GoverningEquations-a}, \eqref{eq:GoverningEquations-b} is the full density from \eqref{eq:GoverningEquationsSupplement-a} and not the dry air density. Therefore, the effect of moisture on (total) density is properly accounted for. Individual contributions from the moist constituents are the following: in the thermodynamic equation \eqref{eq:GoverningEquations-d}, $C$ denotes the ``total moist heat capacity'', specified in \eqref{eq:GoverningEquationsSupplement-i}; $\Sigma$, defined in (2j), collects moist contributions related to the work done by pressure forces and $L(T)$ is the latent heat of vaporization, which can be written as a linear function of temperature under the assumption of constant $c_l$.  The reference values for latent heat and temperature are $L_\text{ref}=2.5\cdot10^6\text{ J/kg/K}$ and $T_\text{ref}=273.15\text{ K}$, respectively. Finally, the right hand side represents temperature changes caused by precipitation. In the transport equations for the respective mixing ratios \eqref{eq:GoverningEquations-e}-\eqref{eq:GoverningEquations-g}, terms of the form $S_\star$ denote the usual Kessler-type closures for the microphysical processes of condensation (cd), evaporation (ev), autoconversion (ac) and collection (cr), respectively. In \eqref{eq:GoverningEquationsSupplement-e}-\eqref{eq:GoverningEquationsSupplement-h}, $C_\star$ denote rate constants of the respective processes, while $q_\text{cn}$ represents the density of condensation kernels and $q_\text{ac}$ denotes an activation threshold for the autoconversion of cloud droplets into raindrops.

We do not explicitly consider cold clouds that would necessitate parametrization of the ice phase. While a comprehensive treatment of cloud formation and precipitation on synoptic scales in the midlatitudes \emph{should} include bulk microphysics closures for the ice phase \citep{houze2014}, we reserve this endeavor for future work.


\section{Overview and comparison of resulting model equations}

For reference, we now briefly summarize and compare the results of the derivations in sections~4-5 (for Smith and Stechmann's anelastic/FARE  PQG model) and section~6 (for the new model) in dimensional form. For brevity, we will refer to the former just as PQG and to the latter as $\text{PQG}_{\text{DL}}$ from here on out. 

Our notation here is as follows: for any model variable $f$ with a leading-order vertical background profile, we denote it by $\bar{f}=\bar{f}(z)$, while $\tilde{f}=\tilde{f}(t,\vec{x},z)$ stands for the corresponding perturbation. We will always assume $\tilde{f}\ll\bar{f}$. Furthermore,  we write
\begin{equation}
D^g_t=\partial_t+\vec{u}\cdot\grad_\parallel
\end{equation}
for the material derivative with respect to the geostrophic horizontal velocity $\vec{u}$.


\subsection{The PQG model}

The following are equivalent to equations (65)-(66) in \citep{smith2017}, if a $\beta-$plane approximation is adopted:

\noindent
\emph{Diagnostic relations:} These are the usual geostrophic and hydrostatic balances,
\begin{subequations}
\begin{align}
\label{eq:Geostrophy}
f\vec{k}\times\vec{u} = -\grad_\parallel{\phi}, 
  \\
\label{eq:Hydrostatics}
g\frac{\tilde{\theta}}{\theta_\text{ref}}=\partial_z\phi,
\end{align}
\end{subequations}
where $\phi=\frac{\tilde{p}}{\bar{\rho}}$ is the pressure perturbation scaled by the background density and $f$ the reference value of the Coriolis parameter.

\noindent
\emph{Transport equations:} As in classical dry air theory, we obtain prognostic equations for the geostrophic vertical vorticity $\zeta$ and for potential temperature; in PQG, we get an additional prognostic equation for total moisture $q_T$:
\begin{subequations}
\begin{align}
\label{eq:PQGVorticityTransportDimensional}
D^g_t\left[\zeta+\beta y\right]&=\frac{f}{\bar{\rho}}\partial_z\left(\bar{\rho}\tilde{w}\right), 
  \\
\label{eq:PQGThermodynamicEquationDimensional}
D^g_t\tilde{\theta}_e+\tilde{w}\derivative{\bar{\theta}_e}{z}&=0, 
  \\
\label{eq:PQGTotalWaterTransportDimensional}
D^g_t\tilde{q}_T+\tilde{w}\derivative{\bar{q}_T}{z}-\frac{1}{\bar{\rho}}\partial_z\left(\bar{\rho}V_r\tilde{q}_r\right)&=0,
\end{align}
\end{subequations}
where $\beta$ denotes the latitudinal variation of the Coriolis parameter, $\zeta=\partial_xv-\partial_yu$ the vertical vorticity and $\tilde{w}$ the small geostrophic vertical velocity; $\theta_e$ denotes the (linearized) equivalent potential temperature, given by
\begin{equation}\label{eq:ThetaEDefinition}
\theta_e=\theta+\frac{L_{\text{ref}}}{c_{\text{pd}}}q_v.
\end{equation}
Owing to the simplified phase changes in the FARE setting \citep{hernandez2013}, $\tilde{q}_r$ and $\tilde{q}_v$ can be written in terms of $q_T$, so that only one equation for moisture is needed (see the next section for details). 

As laid out by \citet{smith2017}, the vertical velocity can be eliminated from this system to yield a potential vorticity formulation. Here we only state the results of this calculation and refer the reader to the cited article for more information. Thus, the potential vorticity $Q$ based on equivalent potential temperature in PQG reads
\begin{equation}\label{eq:PQGPVDefinition}
Q=\zeta+\beta y+\frac{f}{\bar{\rho}}\partial_z\left(\frac{\bar{\rho}\tilde{\theta}_e}{\text{d}\bar{\theta}_e\slash\text{d}z}\right).
\end{equation}
This quantity can be rewritten in terms of the pressure perturbation $\phi$ and the moisture-related variable
\begin{equation}\label{eq:MAndGMDefinitions}
M=\tilde{q}_T+G_M\tilde{\theta}_e,
\qquad
\text{where} 
\qquad
G_M = -\frac{\text{d}\bar{q}_T\slash\text{d}z}{\text{d}\bar{\theta}_e\slash\text{d}z}.
\end{equation}
For given $M$ and $Q$, the result is a second-order diagnostic reconstruction equation for the pressure perturbation, $\phi$, 
\begin{align}\label{eq:PQGInversionEqnRaw}
&\Delta_\parallel\phi+f\beta y+\frac{f^2}{\bar{\rho}}\partial_z\left[\frac{\bar{\rho}}{\text{d}\bar{\theta}_e\slash\text{d}z}H_u\frac{1}{D_M}\left(\frac{\theta_\text{ref}}{g}\partial_z\phi+\frac{L_\text{ref}}{c_\text{pd}}M\right)\right] \nonumber \\
&+\frac{f^2}{\bar{\rho}}\partial_z\left[\frac{\bar{\rho}}{\text{d}\bar{\theta}_e\slash\text{d}z}H_s\left(\frac{\theta_\text{ref}}{g}\partial_z\phi+\frac{L_\text{ref}}{c_\text{pd}}q_\text{vs}\right)\right]=fQ\,,
\end{align}
where $D_M:=1+L_\text{ref}G_M\slash c_\text{pd}$ and $H_u$, $H_s$ are Heaviside switching functions for the unsaturated and saturated state, respectively. 
For later comparison with the PQG$_{\text{Dl}}$ model, and following \citet{wetzel2019}, we rewrite \eqref{eq:PQGInversionEqnRaw} as
\begin{align}\label{eq:PQGInversionEqn}
\Delta_\parallel\phi
+ \frac{1}{\bar{\rho}}
  \partial_z 
  \left[\frac{\bar{\rho}f^2}{N^2}\partial_z\phi \right] 
  = \nonumber \\
f (Q - \beta y) - 
  \frac{f^2}{\bar{\rho}}
  \partial_z 
  \left[\bar{\rho} g L^* \left(H_u \frac{M}{N_u^2} + H_s \frac{q_\text{vs}}{N_s^2} \right)
  \right],
\end{align}
where
\begin{align}\label{eq:SwitchedBVFrequency}
N^2 = H_u N_u^2 + H_s N_s^2
\qquad\text{and}\qquad
L^* = \frac{L_\text{ref}}{c_\text{pd}\theta_\text{ref}}\,.
\end{align}
Here $N$ is the local buoyancy frequency, while $N_u = \sqrt{(g/\theta_\text{ref}) \, \text{d}\bar{\theta}/\text{d}z}$ and $N_s = \sqrt{(g/\theta_\text{ref}) \, \text{d}\bar{\theta}_e/\text{d}z}$ are the buoyancy frequencies in undersaturated and saturated air, respectively.

As Smith and Stechmann point out, the elliptic operator on the left of \eqref{eq:PQGInversionEqn} has non-constant coefficients owing to the switch in the Brunt-Väisälä frequency defined in \eqref{eq:SwitchedBVFrequency}. Moreover, the switching functions depend on $q_T$ and therefore implicitly on $\partial\phi/\partial z$ through the hydrostatic balance in \eqref{eq:Hydrostatics} together with the definitions of $M$ in \eqref{eq:MAndGMDefinitions} and $\theta_e$ in \eqref{eq:ThetaEDefinition}. As a consequence, the elliptic PDE \eqref{eq:PQGInversionEqn} is nonlinear, and its right hand side involves a surface $\delta$-distribution along the phase boundary because the vertical derivative is applied to a discontinuous function. As discussed by \citet{wetzel2019}, the numerical solution of the $PV$-$M$-inversion problem is, for these reasons, much more involved than it is for the classical QG  model.

The transport equation for the PQG potential vorticity reads
\begin{equation}
D_t^gQ=-\frac{f}{\text{d}\bar{\theta}_e\slash\text{d}z}\partial_z\vec{u}\cdot\grad_\parallel\tilde{\theta}_e.
\end{equation}
\emph{Remark:} Smith and Stechmann chose $Q$ and $M$ as given quantities for potential vorticity inversion, because $q_T$ does not constitute a balanced quantity in the sense that its evolution equation depends on the vertical velocity \citep{wetzel2019}.


\subsection{The $\text{PQG}_\text{DL}$ model}

Compared to PQG, this model has a fundamentally different aim: as already stated in the introduction, it is meant to connect to the diabatic layer of \citet{klein2022}, which necessitates various changes in the scaling of the moist constituents. In particular, the mixing ratios here do \emph{not} decompose into a vertical background and corresponding perturbations. The quantities $q_j$ ($j=v,c,r$) therefore simply denote the respective leading-order contributions from moist constituents. Moreover, $\text{PQG}_\text{DL}$ incorporates more detailed cloud microphysics, keeping the original number of moisture species in the asymptotic regime.

\noindent\emph{Diagnostic relations:} The geostrophic and hydrostatic balances take the same form as above, see \eqref{eq:Geostrophy} and \eqref{eq:Hydrostatics}. However, the scaling of the terminal rainfall velocity $V_r$ now produces an additional diagnostic relation for $q_r$, which is determined from
\begin{equation}\label{eq:PQGDLqrBalance} 
\partial_zq_r=-\frac{\bar{\rho}}{V_r}\left[S_\text{ag,m}+S_\text{cr}-S_\text{ev}\right].
\end{equation}
Here, $S_\text{ag,m}$ represents an ad hoc parametrization of the generation of raindrops through aggregation and melting, see section 6 for details.

\noindent\emph{Transport equations:} Since $\text{PQG}_\text{DL}$ does not assume fast autoconversion, the cloud water evolution equation is retained in the leading-order equations, increasing the number of prognostic equations by one relative to PQG,
\begin{subequations}\label{eq:PQGDLTransportEquations}
\begin{align}
\label{eq:PQGDLTransportEquations-a}
D^g_t\left[\zeta+\beta y\right]
  & =\frac{f}{\bar{\rho}}\partial_z\left(\bar{\rho}\tilde{w}\right), 
    \\
\label{eq:PQGDLTransportEquations-b}
D^g_t\tilde{\theta}_e+\tilde{w}\derivative{\bar{\theta}}{z}
  & = c_lV_rq_r\frac{R_d}{c_\text{pd}}\partial_z\ln\bar{p}, 
    \\
\label{eq:PQGDLTransportEquations-c}
D^g_tq_v
  & = S_\text{ev}-S_\text{cd}, 
    \\
\label{eq:PQGDLTransportEquations-d}
D^g_tq_c
  & = S_\text{cd}-S_\text{ag,m}-S_\text{cr}.
\end{align}
\end{subequations}
We observe that rain now influences the system's dynamics in two ways: it moistens dry air by evaporating, while producing a cooling effect in the thermodynamic equation even in saturated air. This constitutes the main departure from PQG, which we will illustrate by means of a diagnostic relation for the vertical velocity later on. Moreover, due to the fact that the moisture constituents do not feature a background stratification and that above heights of $3\, \text{km}$ their mixing ratios are roughly by an order of magnitude smaller than assumed  for PQG, the background potential temperature stratification and that of the equivalent potential temperature are identical in the $\text{PQG}_\text{DL}$ regime, i.e., $\bar{\theta}_e \equiv \bar{\theta}$.

Potential vorticity in the $\text{PQG}_\text{DL}$ regime can be defined exactly as in \eqref{eq:PQGPVDefinition}. The absence of a background distribution of moisture, however, removes the vertical velocity from all moist transport equations. Therefore, instead of \eqref{eq:PQGInversionEqn}, in this scenario the diagnostic equation for the pressure perturbation reads:
\begin{align}\label{eq:PQGDLInversionEqn}
\Delta_\parallel\phi
+ \frac{1}{\bar{\rho}}\partial_z\left(\frac{\bar{\rho} f^2}{N^2} \partial_z\phi \right) = \nonumber \\
f (Q - \beta y) 
- \frac{f^2}{\bar{\rho}}\partial_z\left(\frac{\bar{\rho} L^*}{N^2} q_v \right),
\end{align}
where now 
\begin{equation}\label{eq:PQGDLBVFrequency}
N^2 = \frac{g}{\theta_{\text{ref}}} \frac{\text{d}\bar{\theta}}{\text{d}z},
\end{equation}
while $Q$ and $q_v$ can be chosen as given quantities to initiate potential vorticity inversion. In addition, $q_c$ will be needed to determine $q_r$ from \eqref{eq:PQGDLqrBalance} and complete the diagnostic determination of all model variables. Finally, the evolution equation for potential vorticity also takes a different form, due to the nonzero right hand side in \eqref{eq:PQGDLTransportEquations-b}:
\begin{align}
D_t^gQ
= &- \frac{f}{\text{d}\bar{\theta}\slash\text{d}z}\partial_z\vec{u}\cdot\grad_\parallel\tilde{\theta}_e \nonumber \\
&+\frac{f}{\bar{\rho}}\left[\frac{\bar{\rho}c_lV_rq_r\frac{R_d}{c_\text{pd}}\partial_z\ln\bar{p}}{\text{d}\bar{\theta}\slash\text{d}z}\right].
\end{align}
In contrast to the PV inversion equation for PQG in \eqref{eq:PQGInversionEqn}, its PQG$_{\text{DL}}$ version in \eqref{eq:PQGDLInversionEqn} is linear for given $Q$ and $q_v$, and it features smooth coefficients in the linear operator. Moreover, $q_v$ is generally continuous -- though not continuously differentiable -- so that the right hand side in \eqref{eq:PQGDLInversionEqn} involves merely discontinuities at phase boundaries instead of surface Dirac distributions as in PQG. We conclude that numerical PV inversion in PQG$_{\text{DL}}$ is much less of a challenge than it is in PQG. In the envisioned PQG-DL-Ekman triple deck model, much stronger effects of moist processes will be included, but confined to the intermediate diabatic layer (DL).


\section{Non-dimensionalisation and scaling for PQG}

Let us first recap the generic and space-time-scale independent distinguished limit for the traditional dry air model from \citet{klein2010}: Coupling the usual dimensionless parameters to one small parameter $\epsilon$ with a physical magnitude of $\sim\frac{1}{10}$, we choose the following scaling for the Mach, external Froude and Rossby numbers:
\begin{align}
\text{M} 
= \frac{u_{\text{ref}}}{\sqrt{\nicefrac{p_{\text{ref}}}{\rho_{\text{ref}}}}}=\epsilon^{\nicefrac{3}{2}}
&= \frac{u_{\text{ref}}}{\sqrt{gh_{\text{sc}}}}=\text{Fr}_\text{ext}, \nonumber \\
\text{Ro} &= \frac{u_{\text{ref}}}{2\Omega l_{\text{ref}}}=O(\epsilon),
\end{align}
where the horizontal reference velocity is $u_{\text{ref}}\approx10\text{ m s}^{-1}$, the pressure scale height $h_{\text{sc}}\approx10\text{ km}$ and $l_{\text{ref}}$ is of the order of a synoptic length scale, such that
\begin{equation}
\frac{h_{\text{sc}}}{l_{\text{ref}}}=\epsilon^2.
\end{equation}
Notice that the \emph{internal} Froude number $\text{Fr}_\text{int}=\frac{u_\text{ref}}{Nh_\text{sc}}$, with $N$ denoting the buoyancy frequency, is of the same order of magnitude as $\text{Ro}$, in accordance with classical QG scaling. The buoyancy frequency in this context will depend on the moisture content, see the appendix of \citet{smith2017}. 

We work with the standard $\beta-$plane approximation, within which the Coriolis parameter is a linear function of latitudinal distance $y$:
\begin{equation}
\vec{\Omega}=(f_0+\beta y)\vec{k}.
\end{equation}
As in the textbook derivation of classical QG \citep{pedlosky1987}, the $\beta-$effect is of the order $O(\epsilon)$ compared to $f_0$. 

Regarding moisture, we start by incorporating the simplifications of the FARE model. As described by \citet{hernandez2013}, fast autoconversion and rain evaporation imply that the cloud phase can be omitted altogether, since cloud droplets grow to critical size quasi-instantaneously in this regime, \emph{and} since rain can be recovered from the total water content $q_T=q_v+q_r$ by comparison with the saturation mixing ratio $q_{\text{vs}}$. We can now combine all moisture balances into one by adding the corresponding equations and obtain
\begin{equation}\label{eq:PQG-qr-eqn}
D_tq_T-\frac{1}{\rho_d}\partial_z(\rho_dV_rq_r)=0,
\end{equation}
where
\begin{equation}
q_r=(q_T-q_{\text{vs}})^+.
\end{equation}

The average water vapor content in the mid-latitude troposphere does not exceed a few percent and we assume
\begin{equation}
q_{\text{vs}}=O(\epsilon^2),
\end{equation}
which imposes the same upper bound on the magnitudes of $q_v$, $q_r$. Following \citet{smith2017}, we assume $q_\text{vs}=q_\text{vs}(z)$ to be a given function of altitude. We take the terminal velocity $V_r$ to be constant and, in accordance with \citet{smith2017}, comparable to the vertical reference velocity $w_{\text{ref}}=\epsilon^2u_{\text{ref}}$. This scaling is related to the horizontal reference velocity through the horizontal-to-vertical aspect ratio (the vertical velocity scaling directly follows the classical derivation of the QG model, e.g., in \citep{pedlosky1987}). Thus we let
\begin{equation}
\frac{V_r}{w_{\text{ref}}}=:V_T=O(1).
\end{equation}
These scaling choices were all explicit in Smith and Stechmann's derivation. For the remaining nondimensional parameters, which govern the thermodynamics of moist air, we present a self-contained justification, based on the following guidelines (which we will more comprehensively refer to in the derivation of our new model in section~6):
\begin{enumerate}
\item The ``dry air limit'' that we obtain by letting moisture $q_T\rightarrow0$ in our model equations must agree with the dry QG system.
\item The leading-order balances resulting from our choice of the distinguished limit should encapsulate physically meaningful and - at least qualitatively - accurate relations.
\item Finally, we aim for a scaling that is formally consistent and assigns realistic values to our system parameters.
\end{enumerate}

Following \citet{hittmeir2018}, we proceed with the following general scaling ansatz,
\begin{align}
\frac{L_{\text{ref}}}{c_{\text{pd}}T_{\text{ref}}}=\frac{L}{\epsilon^{a}}, 
\quad
\frac{R_d}{c_{\text{pd}}}&=\epsilon^b\Gamma, 
\quad 
\frac{R_v}{c_{\text{pv}}}=\epsilon^{b_v}\Gamma_v,  
\nonumber \\
\frac{c_l}{c_{\text{pd}}}&=\frac{k_l}{\epsilon^{b_l}},
\quad 
\frac{R_d}{R_v}=\epsilon^cE,
\end{align}
where all parameters on the right-hand sides are $O(1)$ as $\epsilon\rightarrow0$. Thus, $L$, $\Gamma$ etc.\  are \emph{defined} by the above relations, dependent on nonnegative numbers $a,b,\dots$ that are to be determined in accordance with the principles stated above. 

Staying true to guideline~3 and comparing with the physical reference values
\begin{equation}
\frac{L_{\text{ref}}}{c_{\text{pd}}T_{\text{ref}}} \approx 9.1
\qquad\text{and}\qquad
\frac{R_d}{R_v} \approx 0.62
\end{equation}
(cf.~\citet{hittmeir2018}), we are compelled to choose $a=1$ and $c=0$. The choice for $b_l$ is not as obvious, but with the reference value
\begin{equation}
\frac{c_l}{c_{\text{pd}}}\approx4.2
\end{equation}
and $\epsilon\sim\frac{1}{10}$, $b_l=1$ gives the best fit. Moving on to $b$, we have
\begin{equation}
 \frac{R_d}{c_{\text{pd}}}\approx0.29,
\end{equation}
which is quite ambiguous, as far as heuristic matching is concerned, and both $b=0$ and $b=1$ are reasonable choices. A closer look at the temperature equation
\begin{align}
C D_t\ln\theta+\Sigma D_t\ln p+\frac{L(T)}{T}D_tq_v= \nonumber \\
c_lV_rq_r(\partial_z\ln\theta+\frac{R_d}{c_{\text{pd}}}\partial_z\ln p),
\end{align}
however, reveals that - given our scaling up to this point - $b=0$ would raise the rainfall term $c_lV_rq_r\frac{R_d}{c_{\text{pd}}}\partial_z\ln p$ on the right-hand side to a leading-order effect. Yet, no term that depends on $q_r$ occurs in the temperature evolution equation of PQG and this is accomodated here by letting $b=1$. This amounts to the so-called \emph{Newtonian limit} for dry air \citep{ParkinsEtAl2000} (see also the discussion of anelastic models and eqn.~(2.18) in \citep{Bannon1995}). Finally, $b_v$ does not play a part in the leading-order dynamics, and since the derived quantity
\begin{equation}
k_v := \frac{c_{\text{pv}}}{c_{\text{pd}}}\approx 1.8 \sim O(1),
\end{equation}
we choose $b_v=1$ purely for consistency.

With these scalings in place, the dimensionless equations now read
\begin{subequations}\label{eq:GoverningEquationsDimless}
\begin{align}
\label{eq:GoverningEquationsDimless-a}
D_t\vec{u}+\frac{1}{\epsilon}f\vec{k}\times\vec{u}+\frac{1}{\epsilon^3}\frac{1}{\rho}\grad_\parallel{p}
  & = 0, 
    \\
\label{eq:GoverningEquationsDimless-b}
D_tw+\frac{1}{\epsilon^5}\frac{1}{\rho}\partial_zp
  & =-\frac{1}{\epsilon^5}, 
    \\
\label{eq:GoverningEquationsDimless-c}
\partial_t\rho_d+\nabla_\parallel\cdot(\rho_d\vec{u})+\partial_z(\rho_dw) 
  & = 0, 
    \\
\label{eq:GoverningEquationsDimless-d}
C_\epsilon D_t\ln\theta+\epsilon^3\Sigma_\epsilon D_t\ln p+\epsilon\frac{L\phi_\epsilon}{T}D_tq_v &= \nonumber \\
\epsilon^2 k_lV_Tq_r(\partial_z\ln\theta+\epsilon\Gamma\partial_z&\ln p) ,
  \\
\label{eq:GoverningEquationsDimless-e}
D_tq_T-\frac{1}{\rho_d}\partial_z(\rho_dV_Tq_r)
  & = 0,
\end{align}
\end{subequations}
where
\begin{subequations}\label{eq:GoverningEquationsSupplementDimless}
\begin{align}
\label{eq:GoverningEquationsSupplementDimless-a}
\rho
  & = \rho_d(1+\epsilon^2(q_v+q_r)), 
    \\
\label{eq:GoverningEquationsSupplementDimless-b}
p
  & = \rho_dT\left(1+\epsilon^2\frac{q_v}{E}\right), 
    \\
\label{eq:GoverningEquationsSupplementDimless-c}
T
  & = \theta p^{\epsilon\Gamma}\equiv\theta\pi, 
    \\
\label{eq:GoverningEquationsSupplementDimless-d}
C_\epsilon
  & = 1+\epsilon(\epsilon k_vq_v+k_l(q_r)), 
    \\
\label{eq:GoverningEquationsSupplementDimless-e}
\Sigma_\epsilon 
  & = \Gamma k_l(q_r)+\epsilon\kappa_vq_v, 
    \\
\label{eq:GoverningEquationsSupplementDimless-f}
\kappa_v
  & = \left(\frac{c_{\text{pv}}}{c_{\text{pd}}}\Gamma-\frac{1}{E}\Gamma\right), 
    \\
\label{eq:GoverningEquationsSupplementDimless-g}
\phi_\epsilon
  & = 1-\frac{k_l-\epsilon k_v}{L}(T-1), 
    \\
\label{eq:GoverningEquationsSupplementDimless-h}
q_r
  & =(q_T-q_{\text{vs}})^+.
\end{align}
\end{subequations}


\section{Asymptotic expansion and derivation of PQG}

In this section, we derive the anelastic version of PQG given by equations (65)-(66) in \citep{smith2017}. As in QG theory for dry air, we presume an expansion for density, pressure and temperature that admits purely vertical profiles up to $O(\epsilon)$, and the latter are indicated by \emph{single} subscripts. Instead, when the background distribution of some variable at some given asymptotic order is constant, then this is indicated by a \emph{double} subscript. Superscripts in brackets denote perturbation variables with generic dependence on $(t,x,y,z)$. Thus, the expansions, including the velocity components, for the PQG regime read 
\begin{subequations}\label{eq:PQGDynamicVariablesExpansion}
\begin{align}
p
  & = p_0 + \epsilon p_1 + \epsilon^2 p^{(2)} + o(\epsilon^2), 
    \\
\rho
  & = \rho_0 + \epsilon\rho_1 + \epsilon^2\rho^{(2)} + o(\epsilon^2), 
    \\
\theta
  & = \theta_{00} + \epsilon\theta_1 + \epsilon^2\theta^{(2)} + o(\epsilon^2), 
    \\
T
  & = T_{00} + \epsilon T_1 + \epsilon^2 T^{(2)} + o(\epsilon^2), 
    \\
\vecu 
  & = \vecu\ordr{0} + \epsilon\vecu\ordr{1} + o(\epsilon),
    \\
w 
  & = w\ordr{0} + \epsilon w\ordr{1} + o(\epsilon). 
\end{align}
\end{subequations}
The constant leading-order terms in the expansions of both $\theta$ and $T$ are a consequence of  the Newtonian limit, see \citep{klein2006,hittmeir2018}.

One important feature of Smith and Stechmann's PQG model is that it assumes a vertical background profile for water vapor while rain is comparatively weak and of the order of a perturbation of the total water content only. The expansions of the respective mixing ratios then read
\begin{subequations}\label{eq:PQGMoistVariableExpansion}
\begin{align}
q_v
  & = \epsilon^2q_{v0}+\epsilon^3q_v^{(1)}+o(\epsilon^3)\,, 
    \\
q_r
  & = \epsilon^3q_r^{(1)}+o(\epsilon^3)\,,
\end{align}
\end{subequations}
implying \emph{small horizontal and temporal variations} of the total water content. Informally speaking, this condition  will be met as long as the air in our model atmosphere does not dry out too much. In their PQG model, Smith and Stechmann consider a regime in which the moisture perturbation variables determine the local saturation status. To realize this within the asymptotic framework based on an expansion of the saturation water vapor profile via
\begin{equation}\label{eq:PQGqvsExpansion}
q_{\text{vs}}=\epsilon^2q_{\text{vs}0}+\epsilon^3q_{\text{vs}1}+\dots,
\end{equation}
we must assume an atmosphere close to saturation, so that 
\begin{equation}
q_{v0}=q_{T0}=q_{\text{vs}0}\,.
\end{equation}
Then, saturation is reached whereever
\begin{equation}\label{eq:PQGSaturationCondition}
q_T^{(1)}\geq q_{\text{vs}1}\,.
\end{equation}

The analysis of the leading-order momentum and mass balances now proceeds along the same lines as for dry air with the following results:  \\[0pt]

\noindent
\emph{Mass balance:} By the usual reasoning, anticipating \eqref{eq:PQGGeostrophicBalance}, we conclude that
\begin{equation}
w^{(0)}=0
\end{equation}
from the balance at leading order, while the \emph{anelastic constraint}
\begin{equation}\label{eq:PQGMassBalanceOne}
\rho_0\nabla_\parallel\cdot\vec{u}^{(1)}+\partial_z(\rho_0w^{(1)})=0
\end{equation}
holds for the first-order perturbations. This contrasts with derivations using the Boussinesq approximation, where $(\vec{u}^{(1)},w^{(1)})$ would obey an incompressibility constraint.\\[0pt]

\noindent
\emph{Horizontal momentum balance:} At leading order, we obtain geostrophic balance:
\begin{equation}\label{eq:PQGGeostrophicBalance}
f_0\vec{k}\times\vec{u}^{(0)}+\frac{1}{\rho_0}\grad_\parallel{p}=0\,,
\end{equation}
while the first-order perturbation reads
\begin{align}
D_t^{(0)}\vec{u}^{(0)}&+\beta y\vec{k}\times\vec{u}^{(0)}+f_0\vec{k}\times\vec{u}^{(1)} \nonumber \\
&+\frac{1}{\rho_0}\grad_\parallel{p^{(3)}}-\frac{\rho_1}{\rho_0^2}\grad_\parallel{p^{(2)}}=0
\end{align}
with 
\begin{equation}
D_t^{(0)}=\partial_t+\vec{u}^{(0)}\cdot\grad_\parallel 
\end{equation}
since $w^{(0)}\equiv0$. Taking the vertical component of the curl yields
\begin{equation}\label{eq:PQGVorticityTransport}
D_t^{(0)}\left[\zeta^{(0)}+\beta y\right]+f_0\nabla_\parallel\cdot\vec{u}^{(1)}=0\,,
\end{equation}
which is the equation for transport of vorticity in the quasi-geostrophic regime.\\[0pt]

\noindent
\emph{Vertical momentum balance:} The leading-order equation accepts the explicit solutions
\begin{equation}\label{eq:PQGDLHydrostatics}
T_{00}\rho_0(z)=p_0(z)= p_0(0) e^{-\frac{z}{T_{00}}}\,,
\end{equation}
where we have used the ideal gas law to obtain $\rho_0=\frac{p_0}{T_{00}}$, and the leading-order vertical momentum equation to derive the hydrostatic balance relation, $\partial_zp_0=-\rho_0=-\frac{1}{T_{00}}p_0$. At first order we obtain
\begin{equation}\label{eq:PQGDLHydrostaticsOne}
\partial_z p_1=-\rho_1\,,
\end{equation}
which, by application of the ideal gas law, further implies
\begin{equation}
\frac{T_1}{T_{00}}=\partial_z\left(\frac{p_1}{\rho_0}\right).
\end{equation}
For the second-order perturbations, we get the analogous result
\begin{equation}\label{eq:PQGHydrostaticsTwo}
\partial_zp^{(2)}=-\rho^{(2)},
\end{equation}
which we will analyse in more detail shortly. 

In summary, we have derived \eqref{eq:PQGGeostrophicBalance} as the nondimensional form of \eqref{eq:Geostrophy}, while the substitution of \eqref{eq:PQGMassBalanceOne} into \eqref{eq:PQGVorticityTransport} yields \eqref{eq:PQGVorticityTransportDimensional}. \\[0pt]

\noindent
\emph{Moisture balances:} Here, we only need to consider one balance equation for the total water content,
\begin{equation}\label{eq:TotalWaterTransport}
D_tq_T-\frac{1}{\rho_d}\partial_z(\rho_dV_Tq_r)=0\,.
\end{equation}
With the expansion scheme summarized in \eqref{eq:PQGMoistVariableExpansion}--\eqref{eq:PQGSaturationCondition}, the leading-order balance from \eqref{eq:TotalWaterTransport} gives
\begin{equation}\label{eq:PQGTotalWaterTransportOne}
D^{(0)}_tq^{(1)}_T+w^{(1)}\derivative{q_{T0}}{z}-\frac{1}{\rho_0}\partial_z(\rho_0V_Tq^{(1)}_r)=0\,.
\end{equation}
This is the nondimensional equivalent of equation \eqref{eq:PQGTotalWaterTransportDimensional}.\\[0pt]

\noindent
\emph{Temperature transport:} The leading-order balance of \eqref{eq:GoverningEquationsDimless-d} reads
\begin{equation}
D_t^{(0)}\theta^{(2)}+LD_t^{(0)}q_v^{(1)}+w^{(1)}\derivative{\theta_1}{z}+Lw^{(1)}\derivative{q_{v0}}{z} = 0\,.
\end{equation}
Utilizing the (linearised) equivalent potential temperature, as defined in \eqref{eq:ThetaEDefinition}, we can rewrite the above as
\begin{equation}
D_t^{(0)}\theta_e^{(2)}+w^{(1)}\derivative{\theta_{e1}}{z} = 0,
\end{equation}
which is equivalent to equation \eqref{eq:PQGThermodynamicEquationDimensional} in nondimensional form.\\[0pt]

\noindent
\emph{Buoyancy:} We now take a closer look at equation \eqref{eq:PQGHydrostaticsTwo}. By the ideal gas law \eqref{eq:GoverningEquationsSupplementDimless-b}, we get
\begin{equation}
p^{(2)}=\rho_0T^{(2)}+\rho_1T_1+\rho^{(2)}T_{00}+\rho_0T_{00}\left(\frac{1}{E}-1\right)q_{v0},
\end{equation}
which implies
\begin{align}\label{eq:PQGHydrostaticsSecondOrderIntermediate}
&\partial_z\left(\frac{p^{(2)}}{\rho_0}\right)=\frac{\partial_zp^{(2)}+\frac{1}{T_{00}}p^{(2)}}{\rho_0} =\frac{1}{T_{00}}\frac{1}{\rho_0} \nonumber \\
&\left[-T_{00}\rho^{(2)}+\rho_0T^{(2)}+\rho_1T_1+\rho^{(2)}T_{00}+\rho_0T_{00}\left(\frac{1}{E}-1\right)q_{v0}\right] \nonumber \\
&=\frac{T^{(2)}}{T_{00}}+\frac{\rho_1}{\rho_0}T_1+\left(\frac{1}{E}-1\right)q_{v0}.
\end{align}
With the help of \eqref{eq:GoverningEquationsSupplementDimless-c}, we can further rewrite \eqref{eq:PQGHydrostaticsSecondOrderIntermediate} in the form
\begin{equation}
\partial_z\left(\frac{p^{(2)}}{\rho_0}\right)=\frac{\theta^{(2)}}{\theta_{00}}+\pi_1\theta_1+\pi_2+\frac{\rho_1}{\rho_0}T_1+\left(\frac{1}{E}-1\right)q_{v0}.
\end{equation}
Now, the \emph{leading-order} buoyancy, as defined in \citep{smith2017}, reduces to exactly the leading-order (dry) potential temperature perturbation, that is $\frac{\theta^{(2)}}{\theta_{00}}$. Our result here shows quite a few extra terms - all background profiles, however, which means that we can  recover the exact form of the anelastic PQG equations at leading order by shifting them over to the left hand side and then identifying the appropriate definition of the pressure perturbation: setting
\begin{align}\label{eq:PQGPhiDefinition}
\phi:=&\frac{p^{(2)}}{\rho_0} \nonumber \\
&-\int_{0}^{z}\left[\pi_1\theta_1+\pi_2+\frac{\rho_1}{\rho_0}T_1+\left(\frac{1}{E}-1\right)q_{v0}\right]d\zeta,
\end{align}
we recover \eqref{eq:Hydrostatics} in nondimensional form. Since all extra terms are functions of $z$ only, geostrophic balance remains unaffected and our derivation from the compressible equations with refined moist thermodynamics can be viewed as complete, at least at leading order. 

Still, the question remains: where does the initial discrepancy in the form of the moist buoyancy stem from? Essentially, this is caused by what we could call the higher ``asymptotic resolution'' of our expansion ansatz, where the background stratification of pressure, density and temperature is comprised of \emph{two terms of different order} each instead of one. Appendix A1 explains this in more detail by describing a sample derivation of the moist anelastic buoyancy.


\section{Consequences of a different scaling approach: PQG and the diabatic layer}

In section 2, we laid out the guiding principles for our choice of a distinguished limit and applied them to the thermodynamic parameters of moist air. Some of our scaling choices, however, are only valid under quite specific conditions. Motivated by the recent introduction of the novel intermediary \emph{diabatic layer} in \citep{klein2022}, we therefore propose a different distinguished limit that is better suited to connect with this adjacent boundary layer. We then sketch the resulting model, which we will develop in more detail in an upcoming paper.


\subsection{The framework}

First, let us clarify the scope and limitations of our model, which are summarized in the following points:

\begin{enumerate}
\item As stated above, our QG model is meant to be incorporated into a triple-deck boundary layer theory, coupled to the diabatic layer, as well as to the Ekman layer. In particular, this has the consequence that \emph{our scaling assumptions have to hold strictly at altitudes $\geq3\text{ km}$ only, letting certain model variables attain asymptotically larger or smaller values in the lower layers.}
\item On the length - and timescales of QG theory, we cannot capture convective processes explicitly, since they require much higher vertical velocities and a drastically different aspect ratio. Therefore, the primary form of precipitation that we consider here is \emph{stratiform rain,} which is generated from nimbostratus, frequently with cold seeder cells aloft - cf. chapter 6 of \citep{houze2014}.
\item We are interested in scenarios with a very thick cloud cover. Therefore, we take the \emph{highest observed liquid water contents} in stratiform clouds as a baseline for our scaling; we discuss the consequences for our scaling in detail further down.
\item Since we have not incorporated the ice phase and the associated phase changes as of yet, we stick with the simplified subdivision of water into vapor, cloud droplets and rain. However, most stratiform rain actually originates from snow and other ice particles that, while falling from the upper troposphere through the cloud, increase in size and eventually melt. As laid out in \citep{houze2014} it is this two-stage process of aggregation and melting that dominates in stratiform rain, \emph{not} accretion, which is the driving force in convective rain. For the time being, we will therefore - inaccurately - treat liquid cloud water and cloud ice as one substance, assume that all solid hydrometeors eventually turn into rain and parametrize aggregation and melting in the same manner as autoconversion. All this, of course, is subject to future refinement.
\end{enumerate}

Keeping the above in mind, let us discuss our scaling choices in depth:


\subsection{Terminal rainfall velocity}

The terminal velocity $V_r$ of raindrops is strongly dependent on their diameter. In the context of a bulk paramerization and large-scale dynamics, however, the subtleties of the size spectrum cannot be captured. Thus, we opt for the most straightforward approach and assign a constant value to $V_r$, to be understood as an average over the size distribution of raindrops. Further, it is only the relative magnitude $\frac{V_r}{w_{\text{ref}}}=V_T$ that matters for scaling purposes. Next, we therefore consider typical sedimentation velocities in stratiform rain. \\
Large raindrops can approach fall speeds up to $10\text{ m s}^{-1}$, while the average value is considerably lower and also depends on the form of precipitation under consideration (see \citet{khvorostyanov2014}, chapter 12). In convective rain, where also hailstones with much higher fall velocities can occur, the choice $V_T\sim\epsilon^{-2}$ would certainly be appropriate. Stratiform rain, generally speaking, is gentler and does not produce comparably large hydrometeors, leading us to choose
\begin{equation}
V_T\sim\epsilon^{-1}
\end{equation}
for the present regime, which corresponds to velocities comparable to $\sim1\text{ m s}^{-1}$.


\subsection{The saturation mixing ratio}

\begin{figure}[h]
\centering
\includegraphics[width=\linewidth]{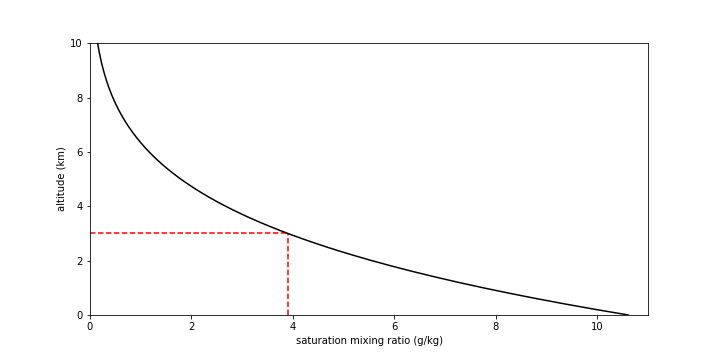}
\caption{Saturation mixing ratio in the mid-latitude troposphere. At 3 km, the top of the diabatic layer, $q_{\text{vs}}$ is already  $<4\text{ g/kg}$, dropping well below $1\text{ g/kg}$ toward the tropopause.}
\label{fig:qvs}
\end{figure}

Sample calculations, using approximate solutions to the Clausius-Clapeyron equation as shown in Fig. \ref{fig:qvs}, show that the saturation mixing ratio $q_{\text{vs}}$ is $\ll10\text{ g kg}^{-1}$ in the bulk of the mid-latitude troposphere (see also sample values in \citep{weischet2018}). We therefore stipulate
\begin{equation}
q_{\text{vs}}\sim\epsilon^3,
\end{equation}
keeping in mind that by asymptotic matching, this still allows the saturation mixing ratio to increase up to $O(\epsilon^2)$ through the diabatic and Ekman layers below. The scaling then agrees very well with the observed magnitudes.


\subsection{Moisture components}

\noindent
\emph{Water vapor:} Supersaturation levels cannot significantly exceed $q_{\text{vs}}$ in magnitude, so that the following scaling for the water vapor content,
\begin{equation}
q_v\sim\epsilon^3,
\end{equation}
suggests itself. \\

\noindent
\emph{Cloud water:} Typically, the cloud water mixing ratio $q_c$ will be smaller than $q_v$ by an order of magnitude, especially in thinner mid-level clouds, as evidenced by observational data, e.g., in \citep{fleishauer2001}. Yet, datasets, e.g., from \citet{zhao2014}, p.96, or \citet{zhang2020} show that the liquid water content in a nimbostratus cloud frequently approaches $0.5\text{ g m}^{-3}$ at altitudes $3$-$4\,\text{ km}$, corresponding to $q_c\sim0.6-0.7\text{ g kg}^{-1}$. With our stated aim to describe an atmosphere with a consistently thick cloud cover, also keeping in mind that we implicitly include cloud ice, this justifies the scaling
\begin{equation}
q_c\sim\epsilon^3\,.
\end{equation}
By letting $q_c^{(0)}\rightarrow0$, we can also discuss ``almost cloud-free'' regions later on, once we have derived our model equations. \\

\noindent
\emph{Rain:} Regarding the rain mixing ratio $q_r$, we simply make the assumption that the total amount of precipitation that eventually reaches the ground is of the same order of magnitude as the total cloud water content. Assuming a water vapor (saturation) mixing ratio of the order $O(\epsilon^2)$ near the ground, this implies $q_r$ should be one order of magnitude smaller than $q_v$. Since nimbostratus clouds extend downward  to rather low levels and rain accumulates as it falls through the various cloud layers, we further expect $q_r$ to increase downwards asymptotically in the same manner as $q_{\text{vs}}$ and $q_v$, indicating that it should be as low as
\begin{equation}
q_r\sim\epsilon^4
\end{equation}
in the free troposphere (3 to 12~km above ground level). Notice that convective processes are filtered out on QG scales, whence exceptionally heavy rain is not covered by our model in absence of parametrizations of the related processes.


\subsection{Phase changes}

Here, we briefly assess which source terms in the  Kessler-type parametrization play a role in the leading-order dynamics. Our guiding principle will be that all microphysical processes which are considered dominant should be represented at leading order in our bulk model as well. This principle is also grounded in the heuristic rule of asymptotic analysis that significant limits should retain a maximal number of terms, known as the \emph{principle of least degeneracy} \citep{kevorkian1996}. In the following, we will argue on purely qualitative grounds and implicitly assume that the respective rate constants scale accordingly; for example, the condensation rate needs to fulfill $C_\text{cd}\slash t_\text{ref}=O(\epsilon^{-3})$ for \eqref{eq:PQGDLScdScaling} to hold. \\

\noindent
\emph{Condensation:} First, let us recall that nimbostratus is primarily generated by gentle updrafts, continually lifting air parcels just above their condensation level. With the QG vertical velocity $w^{(1)}$ being very small, the leading-order equation for $q_c$ will only deal with \emph{horizontal} transport, which might lead one to conclude that condensation does not constitute a leading-order effect in our regime. It should be accounted for, however, that the formation of nimbostratus is significantly fueled by convective cells on the sides of, above or even embedded in the cloud itself (\citep{houze2014}, section 6.4). These processes are external to our model in the sense that they arise from vertical velocities that exceed $w^{(1)}$ by several orders of magnitude. Even though we do not intend to model convection as such, as we have repeatedly emphasized, we argue that the impact of these local events \emph{on the development of the stratiform cloud} suffices to raise condensation to a leading-order effect and therefore opt for the ansatz
\begin{equation}\label{eq:PQGDLScdScaling}
S_{\text{cd}}\sim\epsilon^3
\end{equation}
for the (nondimensionalized) condensation term - which might be supplemented by a parametrization of the aforementioned convective updrafts. \\

\noindent
\emph{Autoconversion, aggregation and melting:} Autoconversion of cloud water by coalescence that causes droplets to cross a size threshold is a very slow process. As already stated above, aggregation and subsequent melting of solid ice particles is the \emph{dominant microphysical mechanism} in stratiform precipitation (\citet{houze2014}, p.\ 144). Hence, we replace $S_{\text{ac}}$ by a source term $S_{\text{ag,m}}$ that parameterizes this mechanism in a manner to be specified later on. Due to our assumption on the maximum cloud water and rain mixing ratios, the scaling
\begin{equation}
S_{\text{ag,m}}\sim\epsilon^3
\end{equation}
is justified. \\[0pt]

\noindent
\emph{Evaporation:} In undersaturated regions, rain evaporates very quickly on QG timescales. In a fully developed nimbostratus cloud, though, it is continually resupplied through melting snow and ice, as described above. The evaporation term should therefore balance $S_{\text{ag,m}}$ in the equation for $q_r$, leading to the asymptotic rescaling
\begin{equation}
S_{\text{ev}}\sim\epsilon^3.
\end{equation}

\noindent
\emph{Accretion:} Here, we refer again to \citet{houze2014}, p.\ 144: while accretion is considered to be the primary growth mechanism in convective clouds, it is of minor importance in nimbostratus. Aiming to account for the presence of convective cells, which might play a significant role in clouds that extend into the diabatic layer, we tentatively include the corresponding source term as well and assume
\begin{equation}\label{eq:PQGDLScrScaling}
S_{\text{cr}}\sim\epsilon^3.
\end{equation}

\noindent
\emph{Remark:} The scaling choices in \eqref{eq:PQGDLScdScaling}-\eqref{eq:PQGDLScrScaling} are physically sensible and necessary to let $\text{PQG}_\text{DL}$ stand on its own. We do not, however, exclude the possibility that connection to the diabatic layer might permit the consideration of different distinguished limits in future work.


\subsection{Thermodynamic parameters}

As far as the scalings of the latent heat and various heat capacities are concerned, we stick with the distinguished limit derived in section 3, with one exception: \\
Since we do not aim to conserve equivalent potential temperature in our variant of PQG, thus allowing precipitation to alter its evolution, the ratio
\begin{equation}
\frac{R_d}{c_{\text{pd}}}=\Gamma
\end{equation}
does not require the ``Newtonian limit'' anymore; we thus assume
\begin{equation}\label{eq:PQGDLNoNewtonianLimit}
\Gamma=O(1)
\end{equation}
from here on out, which yields the standard leading-order solutions for the pressure and density in the troposphere.


\subsection{Asymptotic expansion and derivation}

The governing equations, rescaled for the PQG$_{\text{DL}}$ regime as described above, now read as:
\begin{subequations}
\begin{align}
D_t\vec{u}+\frac{1}{\epsilon}f\vec{k}\times\vec{u}+\frac{1}{\epsilon^3}\frac{1}{\rho}\grad_\parallel{p}&=0, \\
D_tw+\frac{1}{\epsilon^5}\frac{1}{\rho}\partial_zp&=-\frac{1}{\epsilon^5}, \\
\partial_t\rho_d+\nabla_\parallel\cdot(\rho_d\vec{u})+\partial_z(\rho_dw)&=0, \\
C_\epsilon D_t\ln\theta+\epsilon^3\Sigma_\epsilon D_t\ln p+\epsilon^2\frac{L\phi_\epsilon}{T}D_tq_v&= \nonumber \\
\epsilon^2 k_lV_Tq_r(\partial_z\ln\theta+&\Gamma\partial_z\ln p), \\
D_tq_v&= \nonumber \\
S_{\text{ev}}-&S_{\text{cd}}, \\
D_tq_c&= \nonumber \\
S_{\text{cd}}-S_{\text{ag,m}}-&S_{\text{cr}},  \\
\epsilon D_tq_r-\frac{1}{\rho_d}\partial_z(\rho_dV_Tq_r)&= \nonumber \\
S_{\text{ag,m}}+S_{\text{cr}}-&S_{\text{ev}},
\end{align}
\end{subequations}
where
\begin{subequations}
\begin{align}
\rho&=\rho_d(1+\epsilon^3(q_v+q_c+\epsilon q_r))\,, \\
p&=\rho_dT\left(1+\epsilon^3\frac{q_v}{E}\right)\,, \\
T&=\theta p^{\Gamma}\equiv\theta\pi\,, \\
S_{\text{ev}}&=C_{\text{ev}}\frac{p}{\rho}(q_{\text{vs}}-q_v)^+q_r\,, \\
S_{\text{cd}}&=C_{\text{cn}}(q_v-q_{\text{vs}})^+q_{\text{cn}}+C_{\text{cd}}(q_v-q_{\text{vs}})q_c\,, \\
S_{\text{cr}} & = C_{\text{cr}}q_cq_r\,, \\
C_\epsilon&=1+\epsilon^3(k_vq_v+k_l(q_c+q_r))\,, \\
\Sigma_\epsilon&=\Gamma k_l(q_c+q_r)+\epsilon\kappa_vq_v \,, \\
\phi_\epsilon&=1-\frac{k_l-\epsilon k_v}{L}(T-1)\,.
\end{align}
\end{subequations}
The newly introduced source term $S_{\text{ag,m}}$ could be parameterized in analogy with $S_{\text{ac}}$, e.g., as
\begin{equation}
S_{\text{ag,m}}=C_{\text{ag,m}}(q_c-q_{\text{ag,m}})^+,
\end{equation}
where $q_{\text{ag,m}}$ now denotes an activation threshold for the conversion of melting snow into raindrops. Naturally, this is subject to refinement once the ice phase is introduced. 

The scaling assumptions outlined earlier in this section imply changes in the asymptotic expansions relative to those for the PQG regime in \eqref{eq:PQGDynamicVariablesExpansion} and \eqref{eq:PQGMoistVariableExpansion}. Since we do not invoke the Newtonian limit anymore (see \eqref{eq:PQGDLNoNewtonianLimit}), the leading-order temperature is no longer constant, so that 
\begin{subequations}\label{eq:PQGDLTemperatureExpansion}
\begin{align}
T
  & = T_{0} + \epsilon T_1 + \epsilon^2T^{(2)} + o(\epsilon^2) \, .
\end{align}
\end{subequations}
(Recall that $T_{00}$ denoted a constant in the PQG-expansions, whereas here $T_0$ is a $z$-dependent background distribution.) In contrast, the moisture variables expand as
\begin{equation}\label{eq:PQGDLMoistVariableExpansion}
\begin{array}{r@{\ }c@{\ }lc@{\ }l}
\dss q_v
  & =
    & \dss \epsilon^3 q_v\ordr{0} + \dss o(\epsilon^3)\,,
      \\
\dss q_c
  & =
    & \dss \epsilon^3 q_c\ordr{0} + o(\epsilon^3)\,,
      \\
\dss q_r
  & =
    & \dss \epsilon^4 q_r\ordr{1} + o(\epsilon^4)\,.
\end{array}
\end{equation}
That is, their absolute magnitude is reduced by one order in $\epsilon$ relative to the PQG regime (see \eqref{eq:PQGMoistVariableExpansion}), and they do not feature time independent, vertically stratified background states.

With these expansions, the leading-order equations for mass and horizontal momentum remain unchanged relative to those found in the PQG regime. Note, however, that from the second-order vertical momentum balance in \eqref{eq:PQGHydrostaticsTwo}, we may now infer
\begin{equation}
\partial_z\left(\frac{p^{(2)}}{p_0}\right)=\frac{1}{T_0}\left[T^{(2)}+\frac{\rho_1}{\rho_0}T_1\right]
\end{equation}
which involves the leading-order temperature stratification $T_0(z)$. This equation can be rewritten in terms of the potential temperature perturbation to yield the hydrostatic balance \eqref{eq:Hydrostatics}, with a pressure perturbation $\phi$ defined in the same manner as in section~4. 

Let us also recall that by taking the horizontal gradient and utilizing geostrophic balance, we obtain for the potential temperature perturbation $\theta^{(2)}$:
\begin{equation}
\frac{\partial\vec{u}^{(0)}}{\partial z}\cdot\grad_\parallel{\theta^{(2)}}=0,
\end{equation}
which can be viewed as a consequence of the thermal wind relation. 

The explicit solutions for the leading-order pressure and density now read
\begin{equation}
p_0=\left(1-\Gamma z\right)^{\frac{1}{\Gamma}}
\end{equation}
and
\begin{equation}
\rho_0=\left(1-\Gamma z\right)^{\frac{1-\Gamma}{\Gamma}},
\end{equation}
respectively, while the leading-order temperature profile drops off linearly with height:
\begin{equation}
T_0=1-\Gamma z.
\end{equation}

\noindent\emph{Moisture at leading order:} The leading-order equation for water vapor is one for quasi-horizontal transport, since $w^{(0)}=0$:
\begin{equation}
D_t^{(0)}q_v^{(0)}=S_{\text{ev}}^{(0)}-S_{\text{cd}}^{(0)}.
\end{equation}
In the same fashion, we arrive at the equation
\begin{equation}
D_t^{(0)}q_c^{(0)}=S_{\text{cd}}^{(0)}-S_{\text{ag,m}}^{(0)}-S_{\text{cr}}^{(0)}
\end{equation}
for the cloud water content - which is again transported only quasi-horizontally. Cloud formation via condensation is expected to arise through local updrafts in small-scale convective cells, so that this effect is included on the present synoptic scales through an effective parameterization as part of the condensation source term $S_{\text{cd}}^{(0)}$.

For rain, the vertical fallout term dominates:
\begin{equation}\label{eq:PQGDLqrBalanceLeadingOrder}
\partial_z(\rho_0q_r^{(0)})=-\frac{\rho_0}{V_T}\left[S_{\text{ag,m}}^{(0)}+S_{\text{cr}}^{(0)}-S_{\text{ev}}^{(0)}\right],
\end{equation}
while the leading-order potential temperature balance - recalling \eqref{eq:ThetaEDefinition} - reads
\begin{equation}
D_t^{(0)}\theta_e^{(2)}+w^{(1)}\derivative{\theta_{1}}{z}=-k_lV_Tq_r^{(0)}\Gamma\left(1-\Gamma z\right)^{-1}.
\end{equation}
Here we have used that, owing to the systematically lower moisture content compared to the PQG regime, the equivalent potential temperature equals the dry potential temperature to leading and first order. Coupled with the first order mass balance \eqref{eq:PQGMassBalanceOne} and the vorticity transport equation \eqref{eq:PQGVorticityTransport}, this allows us to derive an equation for the QG potential vorticity based on the equivalent potential temperature perturbation $Q=\zeta^{(0)}+\beta y+\frac{f_0}{\rho_0}\partial_z\left(\frac{\rho_0\theta_e^{(2)}}{\text{d}\theta_{1}\slash\text{d}z}\right)$: 
\begin{align}
D_t^{(0)}Q=&\frac{f_0}{\rho_0}\partial_z\left[\rho_0\frac{-k_lV_Tq_r^{(0)}\Gamma\left(1-\Gamma z\right)^{-1}}{\text{d}\theta_{1}\slash\text{d}z}\right]  \nonumber \\
&-\frac{f_0}{\text{d}\theta_{1}\slash\text{d}z}\partial_z\vec{u}\cdot\grad_\parallel{q_v^{(0)}}.
\end{align}
This enables a formulation of our model in terms of potential vorticity inversion, as mentioned in the introduction: given $Q$ and $q_v$, we can rewrite the definition of $Q$ to yield an elliptic partial differential equation for the pressure perturbation, which is given in its dimensional form in \eqref{eq:PQGDLInversionEqn} above. The related PV-inversion problem has a unique solution if we specify boundary conditions for $p$. Horizontal flow and potential temperature can then diagnostically be recovered from the geostrophic and hydrostatic balances, respectively. With given cloud water content $q_c$, we can also use the diagnostic relation \eqref{eq:PQGDLqrBalanceLeadingOrder} to compute the rain water mixing ratio and thus we obtain the leading-order solutions of \emph{all} model variables.


\subsection{Interpretation of the cloud-free limit}

As we already mentioned in our scaling of $q_c$, we will have $q_c^{(0)}\rightarrow0$ quite frequently in a realistic setting (this formal limit simply describes atmospheric environments with a thin cloud cover, it is not meant to imply assumptions on microphysical processes, such as fast autoconversion) - what will be the immediate consequences of this limit for our model? First, we observe that only the evaporation term remains as a sink in the equation for rain:
\begin{equation}
\partial_z(\rho_0q_r^{(0)})=\frac{\rho_0}{V_T}S_{\text{ev}}^{(0)}.
\end{equation}
With the constraint
\begin{equation}
q_r^{(0)}\rightarrow0
\end{equation}
as $z\rightarrow\infty$ and $S_{\text{ev}}^{(0)}\geq0$, this forces us to conclude
\begin{equation}
q_r^{(0)}\equiv0,
\end{equation}
which in turn means that the water vapor mixing ratio is purely advected by the geostrophic flow:
\begin{equation}
D^{(0)}_tq_v^{(0)}=0.
\end{equation}
Thus, \emph{all} moisture components vanish in the leading-order temperature equation. The QG dynamics then becomes that of dry air as expected.


\section{Properties of $\text{PQG}_\text{DL}$: the omega equation}

As we have emphasized repeatedly, $\text{PQG}_\text{DL}$ is set up to develop a more general model, covering synoptic dynamics through the lower troposphere and all the way to the planetary boundary layer. Yet, this model can also stand on its own as a valid extension of etablished QG theory and as a variant of the PQG family of models. Therefore, we will illustrate some of its properties in this section, using the \emph{omega equation.} 

The omega equation is a diagnostic Poisson-type equation for the QG vertical velocity that constitutes a widely used tool in synoptic meteorology \citep{hoskins1978,hoskins1985,hoskins2003}. Since the relationship between moist processes and up- or downdrafts is of general meteorological interest, we will present the omega equation in its $\text{PQG}_\text{DL}-$form and discuss its qualitative properties, also in comparison with PQG. Appendix A2 sketches a derivation of the omega equation in the context of our model.


\subsection{General form of the equation}

As laid out in appendix A2, the omega equation in QG with an arbitrary heat source $S_\theta$ takes the form
\begin{equation}\label{eq:QGOmegaEqn}
N^2\Delta_\parallel\tilde{w}+f^2\frac{1}{\bar{\rho}}\partial_z(\bar{\rho}\partial_z\tilde{w})=2\grad_\parallel\cdot\vec{Q}+f\beta v+\Delta_\parallel S_b,
\end{equation}
where the buoyancy frequency $N=\sqrt{\frac{g}{\theta_\text{ref}}\derivative{\bar{\theta}}{z}}$ is traditionally assumed constant and we have introduced the \emph{Q-vector}
\begin{equation}\label{eq:QVecDef}
\vec{Q}=-\left(\grad_\parallel{\tilde{b}}\cdot\partial_x\vec{u}\right)\vec{i}-\left(\grad_\parallel{\tilde{b}}\cdot\partial_y\vec{u}\right)\vec{j},
\end{equation}
where $\vec{i}$, $\vec{j}$ are unit vectors in the $x$ and $y$ directions, respectively and $\tilde{b}=\frac{g}{\theta_\text{ref}}\tilde{\theta}$ denotes the buoyancy. The second term on the right in \eqref{eq:QGOmegaEqn} is due to the $\beta-$effect, which is frequently neglected. 

As demonstrated in \eqref{eq:AppPQGDLThetaEvolution}-\eqref{eq:AppPQGDLBuoyancyEvolution}, the source term $S_b$ in $\text{PQG}_\text{DL}$ reads
\begin{equation}
S_b=\frac{g}{\theta_\text{ref}}\left[c_lV_rq_r\frac{R_d}{c_\text{pd}}\partial_z\ln\bar{p}-\frac{L_\text{ref}}{c_\text{pd}}(S_\text{ev}-S_\text{cd})\right],
\end{equation}
which leads to
\begin{align}\label{eq:PQGDLOmegaEquation}
N^2\Delta_\parallel\tilde{w}+f^2\frac{1}{\bar{\rho}}\partial_z(\bar{\rho}\partial_z\tilde{w})=2\grad_\parallel\cdot\vec{Q}+f\beta v \nonumber \\
+\frac{g}{\theta_\text{ref}}\Delta_\parallel\left[c_lV_rq_r\frac{R_d}{c_\text{pd}}\partial_z\ln\bar{p}-\frac{L_\text{ref}}{c_\text{pd}}(S_\text{ev}-S_\text{cd})\right]
\end{align}
as the omega equation in $\text{PQG}_\text{DL}$. 

As is apparent from this formulation, the incorporation of cloud microphysics via parametrization of the various phase conversions allows moisture to influence vertical motions in multiple ways:
\begin{itemize}
\item By condensation in saturated air.
\item By ``negative condensation'' (evaporation of cloud water) as well as evaporation of rain in undersaturated air.
\item By falling rain (regardless of satuation), as indicated by the rightmost term in \eqref{eq:PQGDLOmegaEquation}.
\end{itemize}
In PQG variants that only consider one dynamical moisture variable, such as the model derived in section~5, no such variety of interactions is possible: in undersaturated air, moisture cannot influence the dynamics at all.

In the following, we want to make the connection between vertical velocity and the moist constituents more palpable:


\subsection{Estimating moist contributions: a sample solution}

We can isolate the component of vertical velocity that is \emph{directly} due to moist processes by writing any solution of \eqref{eq:PQGDLOmegaEquation} as
\begin{equation}\label{eq:PQGDLWDecomposition}
\tilde{w}=w_d+w_m,
\end{equation}
where $w_d$ solves the ``dry equation'' \eqref{eq:QGOmegaEqn} without any heat sources. Then, $w_m$ will be a solution of
\begin{align}\label{eq:PQGDLOmegaMoist}
N^2\Delta_\parallel w_m+f^2\partial_z\left(\frac{1}{\bar{\rho}}\partial_z\left(\bar{\rho}w_m\right)\right)= \nonumber \\
\frac{g}{\theta_\text{ref}}\Delta_\parallel\left[c_lV_rq_r\frac{R_d}{c_\text{pd}}\partial_z\ln\bar{p}-\frac{L_\text{ref}}{c_\text{pd}}(S_\text{ev}-S_\text{cd})\right].
\end{align}
Let us now prescribe a moisture distribution with a particularly convenient vertical structure that permits an explicit representation of $w_m$ in terms of the sources on the right hand side: in a saturated atmosphere, assume a state of \emph{incipient condensation}, such that $S_\text{ev}=q_r=0$ and
\begin{equation}
S_\text{cd}=C_\text{cn}q_\text{cn}(q_v-q_\text{vs})^+,
\end{equation}
i.e., no significant amount of cloud has emerged yet and all condensation is due to the presence of condensation kernels. We now investigate the case in which $S_\text{cd}$ assumes a vertical structure as follows:
\begin{equation}\label{eq:PQGDLScdScalingII}
S_\text{cd}(t,\vec{x},z)=C_\text{cn}T_0(z)K(t,\vec{x})\sigma(t,\vec{x}),
\end{equation}
where $K$ and $\sigma$ are horizontal distributions of condensation kernels and supersaturation, respectively. Provided such a distribution, we can then assume that $w_m$ also takes such a vertical structure; we therefore look for solutions of the form
\begin{equation}
w_m(t,\vec{x},z)=T_0(z)W_m(t,\vec{x}).
\end{equation}
Substituting this ansatz into \eqref{eq:PQGDLOmegaMoist}, we get for the left hand side
\begin{align}
N^2\Delta_\parallel w_m+f^2\partial_z\left(\frac{1}{\bar{\rho}}\partial_z\left(\bar{\rho}w_m\right)\right)= \nonumber \\
T_0\Delta_\parallel N^2W_m+f_0^2\partial_z\left(\frac{1}{\bar{\rho}}\partial_z\left(\bar{\rho}T_0\right)\right)W_m=T_0\Delta_\parallel N^2W_m,
\end{align}
since $\partial_z(\bar{\rho}T_0)=\partial_z\bar{p}=-\bar{\rho}$. Thus, the vertical component drops out of the equation and, using \eqref{eq:PQGDLScdScalingII}, we are left with
\begin{equation}
T_0\Delta_\parallel N^2W_m=T_0\Delta_\parallel\left[\frac{L_\text{ref}}{c_\text{pd}}C_\text{cn}K\sigma\right],
\end{equation}
which implies
\begin{equation}\label{eq:PQGDLWSpecialSolution}
\Delta_\parallel\left[ N^2W_m-\frac{L_\text{ref}}{c_\text{pd}}C_\text{cn}K\sigma\right]=0.
\end{equation}
Finally, making the assumption that condensation is horizontally restricted to the interior of the domain under consideration, \eqref{eq:PQGDLWSpecialSolution} can be formulated as a Laplace equation with homogeneous Dirichlet boundary conditions \citep{evans2010}, and its unique solution is simply
\begin{equation}
W_m=\frac{L_\text{ref}C_\text{cn}}{c_\text{pd}N^2}K\sigma.
\end{equation}
Thus, if \eqref{eq:PQGDLScdScalingII} holds, updrafts due to condensation intensify \emph{in proportion} with both supersaturation and the concentration of condensation kernels. We emphasize that this result only pertains to updrafts directly related to condensation, not the vertical velocity as a whole.


\section{Conclusions}

In this note, we have demonstrated how to derive the PQG model of \citet{smith2017} in its anelastic form systematically from equations for compressible moist atmospheric flow with bulk microphysics closures. In doing so, we have not introduced a sound-proof limit from the outset, but relied only on the distinguished limit for the Rossby, Mach, and Froude numbers proposed in \citep{klein2010}, and on a heuristic correspondence principle for the asymptotic rescaling of the moist parameters. The derivation was carried out by systematic asymptotic analysis. 

Motivated by the recent introduction of the diabatic layer in the context of QG theory by \citet{klein2022}, we then put forward a modified scaling ansatz, designed to yield a system of equations that can connect to this new intermediate layer (located between the bulk troposphere and the Ekman friction layer) in a mathematically sound and physically meaningful way. In contrast to the PQG model, the new PQG$_{\text{DL}}$ model allows for a straightforward PV inversion, essentially analogous to that of the classical QG theory. In turn,  PQG$_{\text{DL}}$ only models the upper 3 to 12~km of the troposphere, while stronger diabatic effects of moisture at lower levels are to be addressed by the diabatic layer equations, which we intend to work out in a future publication.  Properties of the PQG$_{\text{DL}}$ system were illustrated using the omega equation, an equation that allows to illuminate the relationship between vertical air motions and moist temperature sources. 

We hope to gain more insight into moisture dynamics in the midlatitude atmosphere on the synoptic scale from future investigations of this system, regarding for example baroclinic instability, cyclogenesis and the associated moist processes.

\clearpage

\acknowledgments
D.B. and S.H. acknowledge support by the Austrian Science Fund (FWF) via the SFB ``Taming Complexity in Partial Differential Systems'' with project  number F65. S.H. also acknowledges support by the Austrian Science Fund via the previous Hertha-Firnberg project T-764. R.K. acknowledges support by Deutsche Forschungsgemeinschaft through Grant CRC 1114 ``Scaling Cascades in Complex Systems'', Project Number 235221301, Project (C06) ``Multi-scale structure of atmospheric vortices''. \\
The authors are grateful to the Wolfgang Pauli Institute Vienna for continuous support, e.g. the Pauli fellowship for R.K. and for scientific discussions with its director, N.J. Mauser. \\
The authors thank Leslie Smith and Sam Stechmann for helpful comments, in particular regarding potential vorticity inversion.

\datastatement
The paper presents theoretical work and all derivations should be described in sufficient detail. No further data are required to reproduce the findings.

\appendix


\section{The moist anelastic buoyancy}

In this appendix, we do not aim to derive the full set of anelastic equations; we only want to derive the anelastic buoyancy in the form of \citet{hernandez2013} under a minimal set of scaling assumptions. In this, we will try to stay true to the spirit of the traditional derivation, as it can be found e.g. in \citep{vallis2017}.

\begin{enumerate}
\item The first assumption is that all state variables \emph{as well as the moisture components} can be decomposed into a background state $\tilde{f}=\tilde{f}(z)$ and a perturbation $f'=f'(t,x,y,z)$, where $f'\ll\tilde{f}$. This roughly corresponds to our asymptotic expansion, with the crucial difference that only two terms are considered; however, $f'$ corresponds to $f^{(2)}$ in our regime. Therefore, $f_0$ and $f_1$ here appear ``bundled together'', which leads to a less refined approximation. This is the chief reason for the discrepancy that we encountered in the main text.
\item All thermodynamic relations for the perturbation variables are linearized. This corresponds to only considering the first dynamically relevant term in our asymptotic expansion.
\item The water content relative to dry air in the atmosphere is always small, that is $q_v\ll1$. Furthermore, we consider an unsaturated background state, implying $\tilde{q}_T=\tilde{q}_v$; as an additional consequence, the contribution of water vapor $e$ to the total atmospheric pressure is also small, i.e. $e\ll p$. This does not specify the order of $q_v$ or $e$ relative to the perturbations of the state variables, again in contrast to our regime.
\item The depth of the vertical motion is comparable to the density scale height $-\tilde{\rho}\left(\derivative{\tilde{\rho}}{z}\right)^{(-1)}$; the (potential) temperature scale height is one order of magnitude larger. The first part of this assumption is explicitly stated in \citep{hernandez2013}; the second part makes up for the fact that we do not differentiate between a leading-order \emph{constant} temperature and a first-order background stratification, which again goes back to assumption 1.
\item Moisture is present at leading order and does not drop off too quickly at higher altitudes, that is we require $\frac{\tilde{q_v}}{\tilde{q_{\text{vs}}}}=O(1)$ as well as $\frac{\derivative{\tilde{q_v}}{z}}{\derivative{\tilde{q_{\text{vs}}}}{z}}=O(1)$.
\end{enumerate}


\subsection{Relations for potential temperature}

Since the ideal gas law holds in the form
\begin{equation}
p=R_d\frac{1+\frac{q_v}{E}}{1+q_T}\rho T,
\end{equation}
we can utilize the definition of the potential temperature to derive the relation
\begin{equation}\label{eq:AppTheta}
\ln\theta=\ln C+\frac{1}{\gamma}\ln p-\ln\rho-\ln\bar{R},
\end{equation}
where $C$ is a constant and
\begin{equation}
\bar{R}=R_d\frac{1+\frac{q_v}{E}}{1+q_T}.
\end{equation}
Now, since \eqref{eq:AppTheta} has to hold for the background variables alone, differentiating with respect to $z$ yields
\begin{align}
\frac{1}{\tilde{\theta}}\derivative{\tilde{\theta}}{z}=&\frac{1}{\gamma\tilde{p}}\derivative{\tilde{p}}{z}-\frac{1}{\tilde{\rho}}\derivative{\tilde{\rho}}{z} \nonumber \\
&-\frac{1}{E\left(1+\frac{\tilde{q}_v}{E}\right)}\derivative{\tilde{q}_v}{z}+\frac{1}{1+\tilde{q}_T}\derivative{\tilde{q}_T}{z}.
\end{align}
Since we assume an unsaturated background state, it holds
\begin{equation}
\tilde{q}_T=\tilde{q}_v;
\end{equation}
utilizing hydrostatic balance then yields
\begin{equation}\label{eq:AppBackgroundRelations}
\frac{1}{\tilde{\rho}}\derivative{\tilde{\rho}}{z}=-\frac{g\tilde{\rho}}{\gamma\tilde{p}}-\frac{1}{\tilde{\theta}}\derivative{\tilde{\theta}}{z}-\left(E-1\right)\frac{1}{1+\tilde{q}_v}\derivative{\tilde{q}_v}{z}.
\end{equation}
Now, turning to the perturbations, by linearizing \eqref{eq:AppTheta} we obtain
\begin{equation}\label{eq:AppPerturbationRelations}
\frac{\theta'}{\tilde{\theta}}=\frac{1}{\gamma}\frac{p'}{\tilde{p}}-\frac{\rho'}{\tilde{\rho}}-\frac{q'_v}{E}+q'_T.
\end{equation}


{Derivation of the buoyancy term}

The assumption of hydrostatic balance for the background profiles yields
\begin{equation}
\frac{Dw}{Dt}+\frac{1}{\tilde{\rho}}\frac{\partial\rho'}{\partial z}=-g\frac{\rho'}{\tilde{\rho}}
\end{equation}
in the vertical momentum balance, which can be rewritten as
\begin{equation}
\frac{Dw}{Dt}+\frac{\partial}{\partial z}\left(\frac{p'}{\tilde{\rho}}\right)=-\frac{p'}{\tilde{\rho}^2}\derivative{\tilde{\rho}}{z}-g\frac{\rho'}{\tilde{\rho}};
\end{equation}
substituting \eqref{eq:AppBackgroundRelations}, we then arrive at
\begin{align}\label{eq:AppWEquation}
\frac{Dw}{Dt}+\frac{\partial}{\partial z}\left(\frac{p'}{\tilde{\rho}}\right)= \nonumber \\
-\frac{p'}{\tilde{\rho}}\left[-\frac{g\tilde{\rho}}{\gamma\tilde{p}}-\frac{1}{\tilde{\theta}}\derivative{\tilde{\theta}}{z}-\left(\frac{1}{E}-1\right)\frac{1}{1+\tilde{q}_v}\derivative{\tilde{q}_v}{z}\right]-g\frac{\rho'}{\tilde{\rho}}.
\end{align}
With our assumption on the scale heights, the second term in brackets on the RHS is clearly asymptotically small; the moisture term can be rewritten in the form
\begin{equation}
\frac{1}{\frac{1}{\tilde{q}_v}+1}\frac{1}{\tilde{q}_v}\derivative{\tilde{q}_v}{z}=\frac{1}{\frac{1}{\tilde{q}_v}+1}\derivative{\ln\tilde{q}_v}{z};
\end{equation}
due to $\tilde{q}_v\ll1$, it therefore suffices to show that $\derivative{\ln\tilde{q}_v}{z}$ is not asymptotically large in order to justify the neglect of this term. \\
To this end, we can use the Clausius-Clapeyron equation for the (background) saturation vapor pressure $\tilde{e}_s$:
\begin{equation}
\derivative{\ln\tilde{e}_s}{T}=\frac{1}{R_v}\frac{L(\tilde{T})}{\tilde{T}^2}.
\end{equation}
The saturation vapor pressure, in turn, is related to the (background) saturation mixing ratio $\tilde{q}_{\text{vs}}$ by the formula
\begin{equation}
\tilde{q}_{\text{vs}}=\frac{E\tilde{e}_s}{\tilde{p}-\tilde{e}_s}.
\end{equation}
In our setting, it holds $\derivative{\ln\tilde{q}_v}{z}\sim\derivative{\ln\tilde{q}_{\text{vs}}}{z}$, due to assumption~5. It therefore suffices to estimate the latter:
\begin{align}
\derivative{\ln\tilde{q}_v}{z}=\derivative{\tilde{T}}{z}\derivative{\ln\tilde{e}_s}{\tilde{T}}+\frac{g\tilde{\rho}}{\tilde{p}-\tilde{e}_s}+\frac{\tilde{e}_s}{\tilde{p}-\tilde{e}_s}\derivative{\tilde{T}}{z}\derivative{\ln\tilde{e}_s}{\tilde{T}}= \nonumber \\
\frac{g\tilde{\rho}}{\tilde{p}-\tilde{e}_s}+\left(\frac{\tilde{q}_{\text{vs}}}{E}+1\right)\derivative{\tilde{T}}{z}\derivative{\ln\tilde{e}_s}{\tilde{T}};
\end{align}
the first term here is clearly bounded, while we can employ Clausius-Clapeyron for the second:
\begin{equation}
\derivative{\tilde{T}}{z}\derivative{\ln\tilde{e}_s}{\tilde{T}}=\derivative{\tilde{T}}{z}\frac{1}{R_v}\frac{L(\tilde{T})}{\tilde{T}^2}=\frac{1}{R_v}\frac{L(\tilde{T})}{\tilde{T}}\frac{1}{\tilde{T}}\derivative{\tilde{T}}{z}.
\end{equation}
Due to our assumption on the temperature scale height, we can infer that this term is not just bounded from above, but actually asymptotically small. 

Going back to \eqref{eq:AppWEquation}, the remaining terms yield the expression
\begin{equation}
g\left[\frac{1}{\gamma}\frac{p'}{\tilde{p}}-\frac{\rho'}{\tilde{\rho}}\right],
\end{equation}
which by \eqref{eq:AppPerturbationRelations} can be rewritten as
\begin{equation}
g\left[\frac{\theta'}{\tilde{\theta}}+\left(\frac{1}{E}-1\right)q'_v-q'_c-q'_r\right].
\end{equation}
This is the exact form of the buoyancy force in the anelastic system of \citet{hernandez2013}.


\section{Derivation of the omega equation}

In any version of PQG, the hydrostatic and geostrophic balances (5a) and (5b) hold, as well as the vorticity equation (9a). Since these relations are equivalent to those in standard QG theory and no vertical velocity appears in the moist transport equations of $\text{PQG}_\text{DL}$, any changes in the result will stem from the temperature equation. The latter can be reformulated in terms of potential temperature such that all moist terms appear as sources; thus, it suffices to sketch a derivation that is analogous to one for \emph{dry} QG with a heat source. While the latter can be considered generally known, most references on the omega equation make use of the Boussinesq approximation and neglect the $\beta-$effect. Therefore, we show the main steps of a derivation without these a priori simplifications. 

Our starting point are the geostrophic and hydrostatic balances \eqref{eq:Geostrophy} and  \eqref{eq:Hydrostatics}, respectively, and the transport equations for vorticity and the equivalent potential temperature perturbation \eqref{eq:PQGDLTransportEquations-a} and \eqref{eq:PQGDLTransportEquations-b}, respectively, which we restate here for the reader's convenience:
\begin{subequations}
\begin{align}
\label{eq:AppGeostrophy}
f\vec{k}\times\vec{u}
  & = - \grad_\parallel{\phi}  
    \\
\label{eq:AppHydrostatics}
g\frac{\tilde{\theta}}{\theta_\text{ref}}
  & = \partial_z\phi, 
    \\
\label{eq:AppPQGDLTransportEquations-a}
D^g_t\left[\zeta+\beta y\right]
  & = \frac{f}{\bar{\rho}}\partial_z\left(\bar{\rho}\tilde{w}\right)  
    \\
\label{eq:AppPQGDLTransportEquations-b}
D^g_t\tilde{\theta}_e+\tilde{w}\derivative{\bar{\theta}_e}{z}
  & = c_lV_rq_r\frac{R_d}{c_\text{pd}}\partial_z\ln\bar{p}.
\end{align}
\end{subequations}
As usual, the pressure perturbation $\phi$ assumes the role of a stream function (up to a constant factor), in particular it holds:
\begin{equation}\label{eq:AppVorticityAndStreamfunction}
\zeta=\frac{1}{f}\Delta_\parallel\phi.
\end{equation}
Observing that $\bar{\theta_e}=\bar{\theta}$ in $\text{PQG}_{DL}$, since there is no moist background in this version of PQG and rewriting \eqref{eq:AppPQGDLTransportEquations-b} in terms of $\theta$, we get
\begin{equation}\label{eq:AppPQGDLThetaEvolution}
D^g_t\tilde{\theta}+\tilde{w}\derivative{\bar{\theta}}{z}=c_lV_rq_r\frac{R_d}{c_\text{pd}}\partial_z\ln\bar{p}-\frac{L_\text{ref}}{c_\text{pd}}D^g_tq_v
\end{equation}
and, employing \eqref{eq:PQGDLTransportEquations-c},
\begin{equation}
D^g_t\tilde{\theta}+\tilde{w}\derivative{\bar{\theta}}{z}=c_lV_rq_r\frac{R_d}{c_\text{pd}}\partial_z\ln\bar{p}-\frac{L_\text{ref}}{c_\text{pd}}(S_\text{ev}-S_\text{cd}).
\end{equation}
Finally, assuming a constant Brunt-V\"{a}is\"{a}l\"{a} frequency $N=\sqrt{\frac{g}{\theta_\text{ref}}\derivative{\bar{\theta}}{z}}$, we can write the above in the form
\begin{align}\label{eq:AppPQGDLBuoyancyEvolution}
D^g_t\tilde{b}+N^2\tilde{w}= \nonumber \\
\frac{g}{\theta_\text{ref}}\left[c_lV_rq_r\frac{R_d}{c_\text{pd}}\partial_z\ln\bar{p}-\frac{L_\text{ref}}{c_\text{pd}}(S_\text{ev}-S_\text{cd})\right]=:S_b,
\end{align}
where $\tilde{b}:=g\frac{\tilde{\theta}}{\theta_{\text{ref}}}$. \\
Now, starting with the actual derivation, we can combine \eqref{eq:AppPQGDLTransportEquations-a} and \eqref{eq:AppVorticityAndStreamfunction} to obtain
\begin{equation}
D_t^g(\Delta_\parallel\phi)=\frac{f^2}{\bar{\rho}}\partial_z(\bar{\rho}\tilde{w})-f\beta v.
\end{equation}
Taking the vertical derivative of this equation and using hydrostatic balance \eqref{eq:AppHydrostatics}, we get
\begin{align}\label{eq:AppA24}
D_t^g(\Delta_\parallel\tilde{b})&=-\partial_z\vec{u}\cdot\grad_\parallel(\Delta_\parallel\phi)+\frac{f^2}{\bar{\rho}}\partial_z(\bar{\rho}\tilde{w})-f\beta v \nonumber \\
&=-\vec{k}\times\grad_\parallel{\tilde{b}}\cdot\grad_\parallel{\zeta}+\frac{f^2}{\bar{\rho}}\partial_z(\bar{\rho}\tilde{w})-f\beta v.
\end{align}
In the next step, we take $\Delta_\parallel$ of \eqref{eq:AppPQGDLTransportEquations-b} and get
\begin{align}
D_t^g(\Delta_\parallel\tilde{b})+\Delta_\parallel(\vec{u}\cdot\grad_\parallel{\tilde{b}})-\vec{u}\cdot\grad_\parallel(\Delta_\parallel\tilde{b})= \nonumber \\
-N^2\Delta_\parallel\tilde{w}+\Delta_\parallel S_b.
\end{align}
A laborious, but straightforward calculation then shows
\begin{align}
\Delta_\parallel(\vec{u}\cdot\grad_\parallel{\tilde{b}})= \nonumber \\
\vec{u}\cdot\grad_\parallel(\Delta_\parallel\tilde{b})-\grad_\parallel\tilde{b}\cdot\grad_\parallel\vec{u}-2\grad_\parallel\cdot\vec{Q},
\end{align}
where the Q-vector is defined as in \eqref{eq:QVecDef}. Thus, we obtain
\begin{align}\label{eq:AppA27}
D_t^g(\Delta_\parallel\tilde{b})= \nonumber \\
\grad_\parallel\tilde{b}\cdot\grad_\parallel\vec{u}+2\grad_\parallel\cdot\vec{Q}+\Delta_\parallel S_b-N^2\Delta_\parallel\tilde{w}.
\end{align}
To conclude, we want to combine \eqref{eq:AppA24} and \eqref{eq:AppA27} in order to arrive at a diagnostic equation; checking that it holds
\begin{equation}
-\vec{k}\times\grad_\parallel\tilde{b}\cdot\grad_\parallel\zeta=\grad_\parallel\tilde{b}\cdot\Delta_\parallel\vec{u},
\end{equation}
thanks to the incompressibility of $\vec{u}$, these two terms cancel when substracting \eqref{eq:AppA24} from \eqref{eq:AppA27} and rearranging yields
\begin{equation}
N^2\Delta_\parallel\tilde{w}+f^2\frac{1}{\bar{\rho}}\partial_z(\bar{\rho}\partial_z\tilde{w})=2\grad_\parallel\cdot\vec{Q}+f\beta v+\Delta_\parallel S_b,
\end{equation}
the $\text{PQG}_\text{DL}$ version of the omega equation.

\bibliographystyle{ametsocV6}
\bibliography{references}

\end{document}